APPARENT NON-COEVALITY AMONG THE STARS IN UPPER SCORPIO: RESOLVING THE PROBLEM USING A MODEL OF MAGNETIC INHIBITION OF CONVECTION

James MacDonald and D. J. Mullan

Dept. of Physics and Astronomy, University of Delaware, Newark, DE 19716, USA

ABSTRACT

Two eclipsing binaries in the USco association have recently yielded precise values of masses and radii for 4 low-mass members of the association. Standard evolution models would require these dM4.5 – dM5 stars to have ages which are younger than the ages of more massive stars in the association by factors which appear (in extreme cases) to be as large as ~3. Are the stars in the association therefore non-coeval? We suggest that the answer is No: by incorporating the effects of magnetic inhibition of convective onset, we show that the stars in USco can be restored to coevality provided that the 4 low-mass member stars have vertical surface fields in the range 200 – 700 G. Fields of such magnitude have already been measured on the surface of certain solar-type stars in other young clusters.

1. INTRODUCTION

1.1. Star clusters: the importance of coevality

In astrophysics, the determination of a reliable age for a star is a pre-requisite for reliably interpreting the physical parameters of the star. The principal method for determining the age of a star is to assign it to a cluster for which a Hertzsprung-Russell diagram (HRD) can be constructed, and a main sequence turn-off can be identified. However, this method relies on the fundamental assumption of coevality, i.e. all stars in the cluster have the same age. If an example of a non-coeval cluster could be reliably demonstrated, then the basis for many stellar age determinations could be seriously undermined.

A case in point is the Upper Scorpio association (USco). In recent years, claims of non-coevality among the stars of USco have been published. In this paper, we examine these claims and ask: is it possible to restore the fundamental property of coevality? In the next sub-section, we summarize some of the relevant results which have led a number of research groups to the conclusion of non-coevality.

1.2. The age(s) of stars in USco

From photometry of a sample of mainly B stars, de Greus, de Zeeuw & Lub (1989) determined an age of 5 to 8 Myr for USco, based on stellar evolutionary models of Maeder (1981a, b, c). Preibisch et al. (2002), in addition to the main sequence and post main sequence massive stars, studied a large sample of low mass pre-main sequence stars and found a common age of 5 Myr. More recently, Pecaut et al. (2012) analyzed the F stars in the Upper Scorpius, Upper Centaurus–Lupus and Lower Centaurus–Crux subgroups of the Scorpius–Centaurus OB association. They found the USco F star luminosities to be a factor of ~2.5 *lower* than predicted by four sets of evolutionary tracks for a 5 Myr old population. Since the stars are on pre-main sequence tracks, a lower luminosity corresponds to an older age. To quantify this, Pecaut et al. re-examined the isochronal ages for the USco B, A, and G stars, as well as the M supergiant Antares, and determined a mean age of 11 ± 3 Myr, i.e. older by a factor of ~2 than the common age reported by Preibisch et al. (2002). An even younger age has been proposed by Herczeg &

Hillenbrand (2015): based on comparison with the *β* Pic moving group, they conclude that the low mass stars in USco must have a mean age of ~ 4 Myr. Rizzuto et al. (2016) determined the age and component masses for seven G- to M-type binary systems in USco using the orbital solutions and HST multi-band photometry. They find that their G-type binaries have ages of ~11.5 Myr, consistent with the age estimate of Pecaut et al. (2012). However, once again, Rizzuto et al. find that their M-type binaries are significantly younger than the G-type binaries: the M-type binaries are found to have ages of ~7 Myr. Rizzuto et al note that this age discrepancy corresponds to the following features: the models are found to *under*-predict the luminosity of the M-type stars by 0.8 – 0.15 dex, or equivalently, the models are found to *over*-predict the effective temperature of the M-type stars by 100 – 300 K.

Thus there seems to be an inconsistency between the USco age which is determined using massive stars and the age which is determined using low-mass stars. One interpretation of this age inconsistency might be that a mass-dependent problem (of some kind) exists because of the way in which the ages of stars with different masses were determined. A second possibility for the age inconsistency is that among young stars, radius differences can occur as a result of earlier accretion episodes: such differences could be interpreted erroneously as indicating ages which are too young (Baraffe et al. 2009; Baraffe & Chabrier 2010). A third possibility is to interpret the inconsistency literally, in which case the low-mass stars in USco would be required to have been formed *after* the high-mass stars formed. The third possibility would be hard to understand in terms of star formation: once a massive star forms in an interstellar cloud, the associated ionization and turbulence "brings an end to most star formation" (Herbig 1962). We do not offer a quantitative discussion of any of these three possibilities here. Instead, in order to have guidance in resolving the high-mass/low-mass inconsistency in USco, we consider it worthwhile to consider the case of another young association: the *β* Pic moving group (BPMG).

1.3. The case of BPMG

The moving group associated with the star *β* Pictoris has been the subject of many investigations of its age. The range of ages which various researchers have reported is rather wide (factor of at least 2.5, and possibly as large as 5). For present purposes, it is pertinent to note that the lower end of the age range of BPMG overlaps the upper end of the age range of USco. Therefore, in order to set the stage for the present paper, it is relevant to mention how a consistent interpretation of certain empirical information about stars in the BPMG has emerged in the context of a particular model of magneto-convection.

MacDonald & Mullan (2010) noted that isochronal and Li depletion age estimators give statistically significant differences in the ages of individual stars in the *β* Pic moving group. They proposed that the resolution of this discrepancy lay in the effects of inhibition of convective onset due to the presence of internal magnetic fields in the BPMG stars. The inhibition of convection by internal magnetic fields has the effect of slowing down the process of contraction of pre-main sequence stars towards the main sequence. Since the process of pre-main sequence contraction causes the star to move downward and to the left in the HRD, the slowing down of contraction by magneto-convective processes has the effect that magnetic stellar models with a given age, are found to be larger and cooler than non-magnetic models of stars with the same age.

In the present paper, our goal is to apply our magneto-convective model to interpret the empirical properties of four low-mass stars in USco.

We note that Feiden (2016) has also recently applied a different model of magneto-convection to two of the low-mass stars which we propose to model in USco. We will compare our results with Feiden's in the Discussion (Section 6, below).

## 2. THE IMPORTANT ROLE OF ECLIPSING BINARIES IN MORE STRINGENT TESTING OF MODELS OF STELLAR EVOLUTION

Double line eclipsing binaries provide masses and radii for stars with higher precision than any other method (Torres, Andersen & Gimenez 2010). If the stellar age and composition are also well constrained, these measurements provide stringent tests of stellar evolution models. For a number of short period binaries with M star components, it has been found that standard stellar evolution models predict radii for the stars which are significantly smaller than the empirical radii. That is, empirical radii indicate that the stars are "oversized" ("bloated") relative to the model predictions. Examples among solar neighborhood stars are YY Gem (Torres & Ribas 2002) and CM Dra (Morales et al. 2009). For both systems, the radius excess is at least a $5\sigma$ effect when compared to the (small) statistical errors. A third system is CU Cnc (Ribas 2003) for which evolutionary models underestimate the stellar radii by as much as 10%.

Since all three of the above binaries show evidence for magnetic activity (flares and spots) an attractive explanation for the oversizing is that it is due to the presence of an internal magnetic field. The magnetic field leads to increased radius either because of magnetic inhibition of convective onset (Mullan & MacDonald 2001) or the presence of surface dark spots (Chabrier, Gallardo, & Baraffe 2007) or a combination of the two.

MacDonald & Mullan (2014) applied a combined magnetic inhibition and spot model to the stars in these 3 binary systems and determined that the strength of the vertical component of the magnetic field at the surface of the star ranged from 320 to 560 G. Combining these values with the surface area of the star, MacDonald & Mullan (2014) found values for the magnetic fluxes which were in good agreement with the magnetic fluxes which can be derived from the empirical X-ray luminosity–magnetic flux relation (Pevtsov et al. 2003; Feiden & Chaboyer 2013).

In the 3 short period binaries in the solar neighborhood discussed by MacDonald & Mullan (2014), the stars experience strong dynamo action because of tidal locking of axial rotation to the (short) orbital period. However, in the case of USco, the stars are young enough that magnetic braking has not had enough time to remove significant angular momentum. As a result, the stars in USco are expected to have a distribution of rotation rates and a range of surface magnetic field strengths arising from dynamo action. This leads us to expect that some fraction of the USco cluster members may be "oversized" due to magnetic affects. MacDonald & Mullan (2013) proposed this as an explanation of the finding by Jackson, Jeffries & Maxted (2009) that the mean radii for low-mass M-dwarfs in the (relatively) young, open cluster NGC 2516 (age = 125 ± 25 Myr) are larger than model predictions at a given absolute I magnitude or I − K color and also larger than measured radii of magnetically inactive M-dwarfs.

The reliable identification of red dwarfs with empirical radii which are definitely "oversized" relative to evolutionary predictions has become possible in recent years only because the precision with which stellar radii can be determined has improved to the point that the empirical uncertainties in radii are now better than a few percent. In the best cases, the uncertainty in radius is better than 1%. The latter case applies to two of the USco stars which we will attempt to fit in the present paper: the smallness of the errors in masses and radii impose significant constraints on the permitted ranges of the magnetic parameters.

## 3. PRECISE EMPIRICAL DATA FOR 2 ECLIPSING BINARIES IN USCO

In this section we summarize the empirical properties of each component of the two low-mass binaries which are members of USco. In each of these binaries, the two components turn out to have nearly equal masses, but the stars in one binary are 3 times more massive than the stars in the second binary. In the more massive binary (USco5), the components have masses in the range 0.30 – 0.35 $M_\odot$: such stars are close to the boundary where theoretical models of stellar structure predict that main sequence stars undergo a transition to complete convection. However, in the course of pre-main sequence (and early main sequence) evolution, such stars may contain a radiative zone (RZ) for a more or less extended interval of time. In view of this, and in order to distinguish the USco binaries from each other, we shall refer for brevity to the components in USco5 as RZ stars (but note that at the ages which are relevant to stars in the USco association, the USco5 components are in fact found, according to our models, to be fully convective). On the other hand, the less massive binary (EPIC 203710387) contains components with masses close to 0.1 $M_\odot$: such stars are completely convective (CC) at all stages of evolution. We shall refer to these components as CC stars.

3.1. Two RZ M dwarfs in USco: empirical data

UScoCTIO 5 (USco5) was discovered to be a low-mass spectroscopic binary system with orbital period 34 d by Reiners, Basri & Mohanty (2005). The components are both of spectral type M4.5. Membership of the Upper Sco OB association was first proposed by Ardila et al. (2000) and confirmed by Reiners et al. (2005) based on the detection of Li absorption lines. Kraus et al. (2015) showed that primary and secondary eclipses are apparent in the K2 extended *Kepler* mission light curves. From the existing radial velocity measurements of Reiners et al. (2005) and analysis of the light curve, Kraus et al. (2015) determined the component masses and radii to be $M_A = 0.329 \pm 0.002$ $M_\odot$, $R_A = 0.834 \pm 0.006$ $R_\odot$, $M_B = 0.317 \pm 0.002$ $M_\odot$, and $R_B = 0.810 \pm 0.006$ $R_\odot$. Note that the relative errors in both masses and radii are better than 1%: as a result, obtaining satisfactory fits between models and empirical data is a non-trivial exercise.

Kraus et al. (2015) compared the location of the components in the HRD to a number of standard (non-magnetic) stellar evolutionary models and found that none of the evolutionary models gave satisfactory agreement with the observations. A detailed analysis (Pecaut, Mamajek & Bubar 2012) finds an age for the OB association of 11 ± 3 Myr. At this age, the models predict an effective temperature for the M dwarfs in USco5 of $T_{eff}$ ~ 3400 K: this is higher than that determined from the spectral type, $T_{eff} = 3200 \pm 75$ K, a 2.5$\sigma$ discrepancy. Equivalently, the empirical stellar radii for the M dwarfs in USco5 are found to be greater than the (non-magnetic) model predictions by ~10%.

An independent analysis of the USco5 K2 light curve by David et al. (2016) gives slightly larger masses $M_A = 0.3336 \pm 0.0022$ $M_\odot$, $M_B = 0.3200 \pm 0.0022$ $M_\odot$ but significantly greater radii $R_A = 0.862 \pm 0.012$ $R_\odot$, $R_B = 0.852 \pm 0.013$ $R_\odot$. Note that in these cases, although the masses are still determined to better than 1%, the radii are now determined with somewhat poorer precision. Nevertheless, even the errors in the radii are still not much larger than 1%, and therefore present a challenge to attempted model fits.

3.2. Two CC M dwarfs in USco: empirical data

David et al. (2016) also reported the discovery of three new low-mass double-lined eclipsing binaries in the Upper Scorpius association. Of these three, one contains a G star and a K star, and one contains two brown dwarfs: in the present paper, where we are interested in M dwarfs, we will not discuss these two systems. But the third system (EPIC 203710387), containing nearly identical CC M dwarfs, is of interest to us. David et al. (2016) have determined masses and radii as follows: $M_A = 0.1183 \pm 0.0028$ $M_\odot$, $M_B = 0.1076 \pm 0.0031$ $M_\odot$ and $R_A = 0.417 \pm 0.010$ $R_\odot$, $R_B = 0.450 \pm 0.012$ $R_\odot$. For these stars, the errors in masses and radii are not quite as good as for the components of USco5: nevertheless, with errors in the 2 – 3% range, the data of David et al. (2016) are certainly adequate to offer challenges to obtaining good fits between data and stellar models.

A noteworthy feature of the empirical data reported by David et al. for the CC stars in USco is this: of the two stars in the binary, the *lower* mass star is found to have the *larger* radius. This is reminiscent of the unusual case of another pair of young stars where the lower mass component was observed to be hotter than the higher mass component (Stassun et al. 2007). Standard evolutionary models *cannot* explain the existence of a larger radius in the lower mass component if the two components are coeval. In our view, a possible resolution of this discrepancy is that the lower mass star has a stronger magnetic field than the higher mass component: the presence of a stronger field in the low mass component gives rise to the radius inversion. (An analogous resolution of the "surprising" temperature inversion reported by Stassun et al. [2007] has been reported by MacDonald & Mullan [2009].)

4. CODE AND MODELS

Our code has already been described in MacDonald & Mullan (2012, 2013, 2014). Here we note only some comments which are necessary to set the stage for the present application of the code.

4.1. Mixing length ratio and boundary conditions: three approaches

In order to set our results in context, we start by calculating 3 standard (i.e. non-magnetic) models of the evolution of a star with a given mass using three approaches to the boundary conditions (see Mullan et al. 2015). For all 3 models, we use the SCVH+Z equation of state, which is constructed by adding the contribution from heavy elements to the hydrogen – helium equation of state of Saumon, Chabrier & Van Horn (1995). The 3 models (which we refer to as belonging to Sets A, B, and C) differ in the way we select (a) the mixing length, and (b) the outer boundary condition. For set A, we use a mixing length ratio, $\alpha = 1.7$. This is the value of $\alpha$ which allows our code to obtain models that are best fits to the observed properties of the current Sun. In Set A, the outer boundary conditions on temperature and pressure are determined by applying the Eddington approximation at optical depth 0.1. The modelling of set B is the same as for set A except that we use $\alpha = 1.0$. This choice of mixing length ratio is used for ease of comparison with our third set of models, Set C. For Set C the outer boundary conditions on temperature and pressure are set by their values at the base of BT-Settl atmosphere models (Allard, Homeier & Freytag 2012, Allard et al. 2012, Rajpurohit et al. 2013) at optical depth $10^3$. We use $\alpha = 1.0$ for consistency with the actual value of the mixing length ratio which was used in the calculation of the BT-Settl atmosphere models.

In Figures 1 – 3 below, we will present the results we have obtained for the evolutionary track of radius $R$ and $T_{eff}$ in the pre-main sequence phase of evolution for the RZ stars (USco5). In Figures 4 – 6, we will present analogous tracks for our chosen 2 CC stars in USco. In all cases, the first step in our work will be to calculate standard (non-magnetic) evolutionary models starting from a fully convective Hayashi phase model of high luminosity. To provide guidance for the evolutionary time-scales, each figure will include three isochrones to demonstrate how long it takes for a star to move along a track: the isochrones (which will appear as nearly horizontal lines in Fig. 1 – 6) are plotted for ages in Myr of 8 (upper line), 11, and 14 (lowest line). This choice of ages is made so as to bracket the best estimates of the actual age of the stars in USco.

In Figures 1 – 6, the left-most evolutionary track will indicate the results which we have obtained for non-magnetic models. These can be used as fiducial tracks so that the reader can appreciate the changes that occur when a non-standard (magnetic) model with the same mass is eventually computed. Inspection of

Figures 1 – 6 indicates that the fiducial tracks are in all cases *inconsistent* with the empirical values of *R* and $T_{eff}$.

In order to extend models in Sets A and B to include magnetic effects, our approach is based on a physics-based treatment of magnetic inhibition of convective onset (see Section 4.2 below). The tracks obtained for various magnetic models of this kind are plotted in Figs. 1, 2, 4, and 5. We will find that the larger the field, the cooler the models, i.e. the more the corresponding evolutionary track is displaced towards the *right-hand* side of the plot.

However, in the case of Set C models, we do not model magnetic effects in terms of convective inhibition. Instead, we take a different approach: we model the effects of magnetic fields by imposing dark spots which occupy a certain fraction of the stellar surface. Set C is used to determine the degree of surface spot coverage needed to reconcile measured and calculated stellar surface properties.

4.2. Magnetic inhibition model

Our magnetic inhibition model is based on work of Gough & Tayler (1966: hereafter GT), who derived a criterion for the onset of convective instability in the presence of a magnetic field based on an energy principle which is widely used in plasma physics applications (Bernstein et al. 1958). The original GT criterion depended on a single magnetic inhibition parameter, $\delta$. However, in view of dynamo considerations which cannot be expected to generate a magnetic field that is stronger than a certain limit, when the GT criterion is applied to stars, the introduction of a second parameter is also advisable: an upper limit, or "ceiling", should be imposed on the magnetic field strength, $B_{ceil}$ (MacDonald & Mullan 2012). An important aspect of our "two-parameter' solutions is noteworthy: our magnetoconvective best-fit solutions for any given star are much more sensitive to the value of the parameter $\delta$ than to the value of the second parameter, $B_{ceil}$. This allows us to determine the best fit value of $\delta$ within uncertainties which are no worse than 10's of percent even when the allowed values of $B_{ceil}$ may be spread over several orders of magnitude. The significance of this aspect has been stressed by MacDonald & Mullan (2014): "despite the broad range of uncertainty in the *deep interior* field strengths, this does *not* translate …into an equally broad range of *surface* field strengths…. our predictions of the surface field are subject to observational test, even if the $B_{ceiling}$ values are not".

The magnetic inhibition of convective onset is modelled by modifying the usual Schwarzschild criterion for convective onset to become

$$\nabla > \nabla_{ad} + \Delta, \tag{1}$$

Here, $\nabla$ and $\nabla_{ad}$ are the structural and adiabatic logarithmic temperature gradients with respect to pressure, and

$$\Delta = \frac{\delta}{\theta_e}, \tag{2}$$

where $\theta_e = -\partial \ln \rho / \partial \ln T \big|_P$ is a dimensionless thermal expansion coefficient.

In the original GT criterion, which is derived under the assumption that ideal gas conditions prevail, $\delta$ is related to the vertical component of the magnetic field, $B_V$, and the gas pressure, $P_{gas}$, by

$$\delta = \frac{B_V^2}{B_V^2 + 4\pi\gamma P_{gas}}, \qquad (3)$$

where $\gamma$ is the first adiabatic exponent. The modification in equation (2) compared to the original GT criterion is made in order to take into account non-ideal gas effects (MacDonald & Mullan 2009).

In general, $\delta$ may vary as a function of the radial position in a star. Based on dynamo concepts (MacDonald & Mullan 2012), we take $\delta$ to have a constant numerical value from the surface down to the radial location at which the local magnetic field strength reaches $B_{ceil}$. At deeper radial locations, the field is held fixed at $B_{ceil}$. In the first set of models which we present in what follows, we set $B_{ceil}$ = 1 MG. We note that field strength estimates of order 1 MG have been obtained in the deep interiors of low-mass stars in an independent context, namely, in a rotationally-driven interface dynamo in early M dwarfs (Mullan et al. 2015): however, without further detailed modeling of such a dynamo, it may be premature to claim that the results of Mullan et al. (2015) prove rigorously that deep interior fields of order 1 MG can be achieved in practice in low-mass stars such as the ones we discuss here in USco. We will discuss models with different values of $B_{ceil}$ in section 5.3.

To determine the convective energy flux in our magnetic models, we replace $\nabla_{ad}$ by $\nabla_{ad} + \Delta$ everywhere it appears in the mixing length theory for convection.

4.3. Spot model

Certain low-mass stars are known to have dark spots which are present temporarily on their surface (e.g. Kron 1952). These dark spots, which can have areas as large as 20% of the visible stellar hemisphere (e.g. Bopp & Evans 1973), temporarily block a significant fraction of the outward energy flux which wells up from the deep interior of the star. If enough flux is blocked at the surface, the stellar parameters such as $R$ and $T_{eff}$ are expected to adjust in order to handle the internal redistribution of flux. In our work, the approach to including the blocking effects of surface dark spots is based on work by Spruit & Weiss (1986) and Spruit (1992): see MacDonald & Mullan (2012, 2013) for a description of our approach. In this model, there is a single additional parameter, $f_s$, the effective fraction of the surface covered by spots, which are assumed to be completely dark. The surface boundary condition is then

$$L = 4\pi R^2 \left(1 - f_s\right) \sigma T_u^4, \qquad (4)$$

where $T_u$ is the surface temperature of the immaculate (unspotted) surface, which is set equal to the model temperature at optical depth 2/3.

Note that in modeling the presence of dark spots on the surface of a low-mass star in this way, no assumption whatsoever is made about the operation of magnetic effects in the spots: it is simply ***assumed*** that (for unspecified reasons) a completely dark patch with a certain area exists on the star's surface. In view of this assumption, there is no direct method of using the size of a dark spot on a star to estimate the strength of the magnetic field which might (by analogy with sunspots) be present. Indirect methods might be devised to estimate the field strength (e.g. see Section 5.1 below), but too many unknown parameters are involved to allow reliable estimates.

4.4. Estimation of fit parameters

To determine best fit parameters and their standard deviations, we use a Monte Carlo method. We assume that the mass, radius and $T_{eff}$ determinations are from normal distributions. By interpolation in the model grids, we find the age and e.g. magnetic inhibition parameter values that match best the empirical radius and $T_{eff}$. The mean and standard deviation of a fit parameter are determined from 100,000 trials.

## 5. RESULTS

We have applied our models to the observations of the RZ and CC stars in USco. Since the stellar radii have very small uncertainties, the luminosity and effective temperature are highly correlated. Hence, instead of plotting our results in the traditional HRD, we find it more instructive to use a diagram in which the radius $R$ is plotted against effective temperature $T_{eff}$. Kraus et al. (2015) and David et al. (2016) have both relied on spectral types and flux ratios to determine $T_{eff}$.

5.1 The RZ system in USco

Since the components of USco5 are very similar, instead of plotting the components individually, we have modelled the mean of the two objects as a single track in Figures 1-3.

In Figure 1, we compare the location of the "mean" USco5 in the $R$-$T_{eff}$ diagram with evolutionary tracks of magnetic models from set A for mass, M = 0.32 $M_\odot$. Each magnetic track corresponds to a field with a particular value of $\delta$, ranging from 0 (the standard model: left-most track in Fig. 1) to 0.06 (the right-most track in Fig. 1). The rectangle with the solid border shows the position of the mean component using the results of Kraus et al. (2015). The rectangle with the dashed border shows the position of the mean component using the results reported by David et al. (2016). With the results plotted in Fig. 1, we find that the best fit between our magnetic tracks and the empirical data for the "mean" component occurs for $\delta = 0.039 \pm 0.008$. Also shown in Fig. 1 are isochrones for ages 8, 11 and 14 Myr. Our fitting technique also gives a value for the age of the magnetic model which best fits the empirical $R$ and $T_{eff}$. The observational constraints of Kraus et al. (2015) allow us formally to compute slightly different tracks for the two components of USco5: the best formal fits are obtained for ages of 10.5 ± 0.9 Myr for USco5 A and 11.3 ± 1.0 Myr for USco5 B. For the observational constraints of David et al. (2016), we obtain ages of 9.9 ± 2.0 Myr for USco5 A and 10.6 ± 2.0 Myr for USco5 B. For the A component, $\delta = 0.040 \pm 0.015$ and for the B component, $\delta = 0.042 \pm 0.014$. Clearly, these estimates of ages and best-fit $\delta$ values are so close to each other for the two components of USco5 that there is no statistically significant difference between them. This justifies our use of plotting only a single set of tracks for USco5 A and USco5 B in Fig. 1. The best-fit values of $\delta$ for USco5 (according to models belonging to Set A) correspond to surface values of $B_V$ in the range 320 – 470 G.

Figure 2 is the same as Figure 1 except that the models are from set B. Agreement with the observational constraints of Kraus et al. (2015) are obtained for ages 10.8 ± 0.9 Myr for USco5 A and 11.6 ± 1.0 Myr for USco5 B. For both components, $\delta = 0.030 \pm 0.008$. For the observational constraints of David et al. (2016), we obtain ages 10.2 ± 2.0 Myr for USco5 A and 10.8 ± 2.0 Myr for USco5 B. For the A component, $\delta = 0.031 \pm 0.016$ and for the B component, $\delta = 0.034 \pm 0.015$. As before, we note that there are no statistically significant differences between ages and best-fit $\delta$ values for USco5 A and USco5 B. The predicted range in $B_V$ (according to models belonging to Set B) is 230 – 420 G.

Turning now to models belonging to Set C, we show in Figure 3 a comparison between the location of USco5 in the $R$-$T_{eff}$ diagram and the location of evolutionary models containing dark spots for a range of values of the fractional areal coverage $f_s$. Agreement with the observational constraints of Kraus et al. (2015) are obtained for age 10.0 ± 0.9 Myr for USco5 A and 10.8 ± 1.1 Myr for USco5 B. For both components, the best fit models have $f_s = 0.33 \pm 0.06$. For the observational constraints of David et al.

(2016), we obtain ages of 9.6 ± 2.1 Myr for USco5 A and 10.4 ± 2.2 Myr for USco5 B. For the A component, $f_s = 0.35 ± 0.15$ and for the B component, $f_s = 0.34 ± 0.14$. Once again, there are no statistically significant differences between the results for USco5 A and USco5 B.

In the case of Set C, there is no obvious way to predict the surface magnetic field strength which corresponds to a given fractional areal spot coverage, unless one makes several assumptions. E.g. Bopp & Evans (1973) suggest that the magnetic field strength in a star spot can be calculated if one knows the following information: the thickness of the spherical shell where energy is trapped, the efficiency with which the radiative flux deficit is converted into magnetic energy, and the length of time during which the flux deficit is stored in magnetic energy before being released in a flare. For lack of information concerning these factors in the RZ and CC stars in USco, we do not make any estimates in this paper concerning magnetic field strengths in Set C models.

We draw attention to an important quantitative conclusion that can be inferred from Figs 1-3. Whether we use boundary conditions from Sets A, B, or C, the ages which emerge from our best-fit magnetic models of the RZ stars in USco5 are found to lie in the range 9.6 ± 2.1 to 11.6 ± 1.0 Myr. These results are consistent with the ages obtained for the more massive stars in USco: 11 ± 3 Myr (as described in Section 1.2 above).

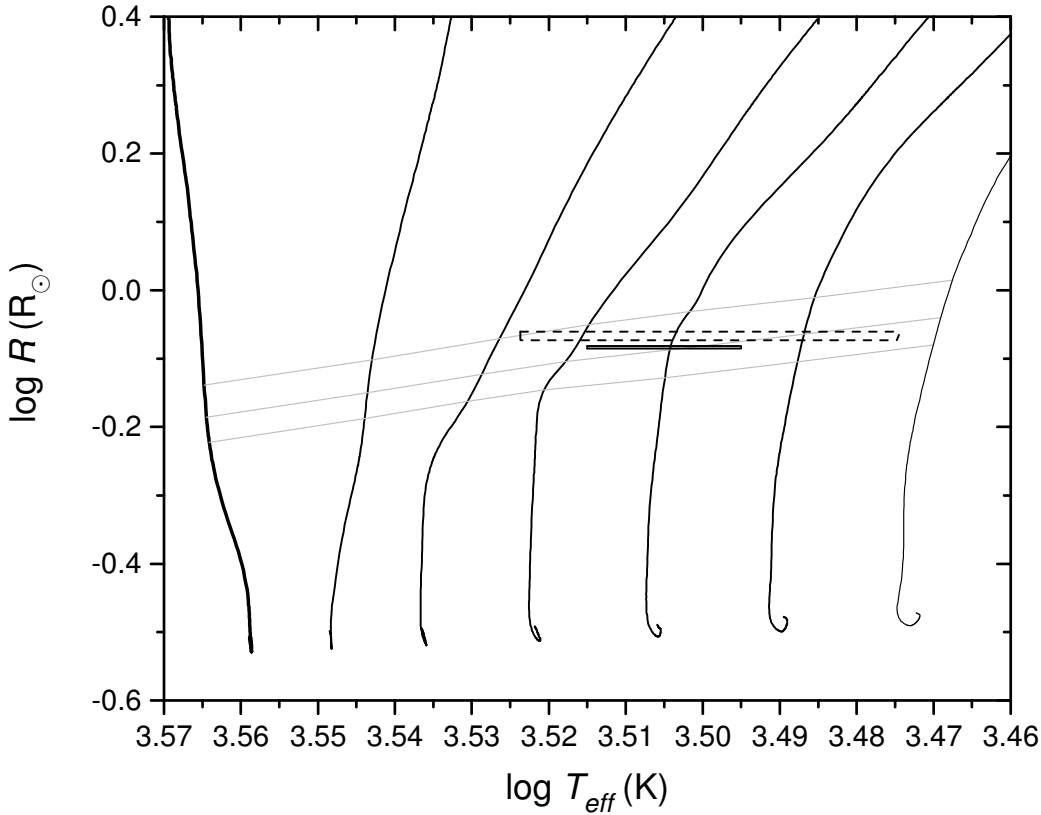

Figure 1. Comparison of the location of RZ system USco5 in the log $T_{eff}$ – log $R$ diagram with models from set A (see Section 4.1) for M = 0.32 $M_\odot$. The magnetic inhibition parameter ranges from 0 (leftmost track: non-magnetic "standard evolution") to 0.06 (rightmost track) in increments of 0.01. The rectangles show the position of the mean component using the results of Kraus et al. (2015, solid border) and David et al. (2016, dashed border). Also shown are isochrones for ages 8 (top), 11 and 14 (bottom) Myr.

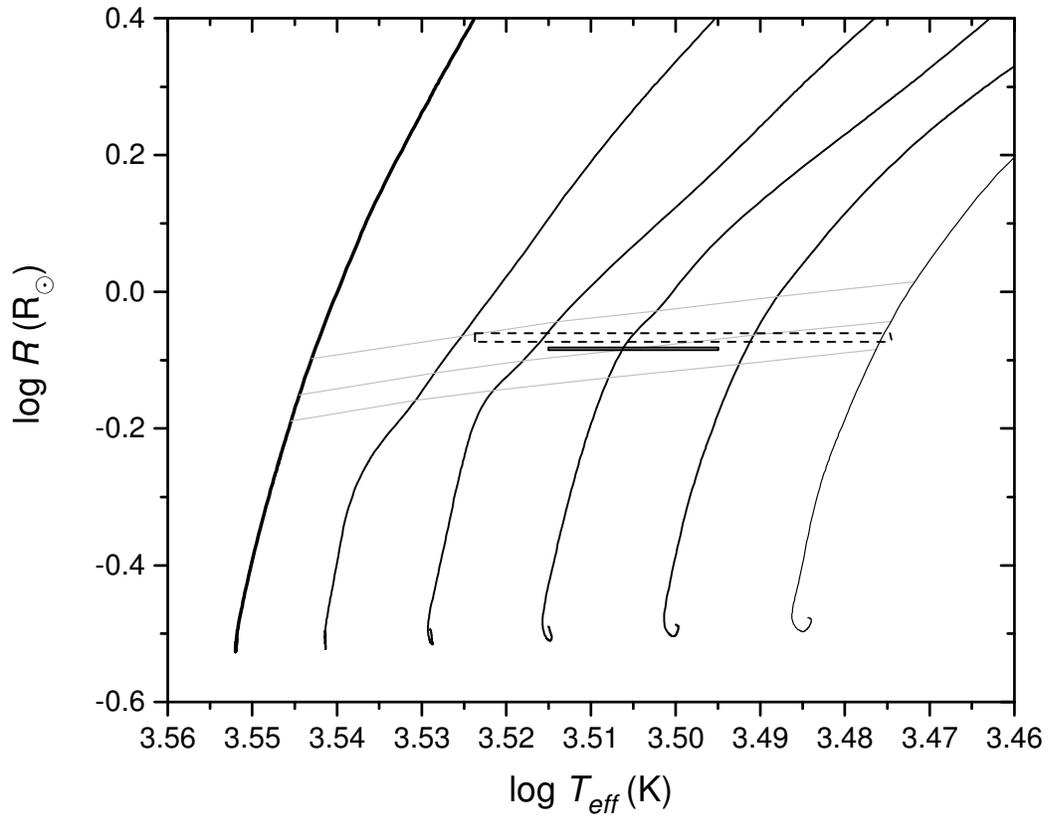

Figure 2. Analogous to Figure 1, except for models from set B (see Section 4.1). Here, the magnetic inhibition parameter ranges from 0 (leftmost track) to 0.05 (rightmost track) in increments of 0.01.

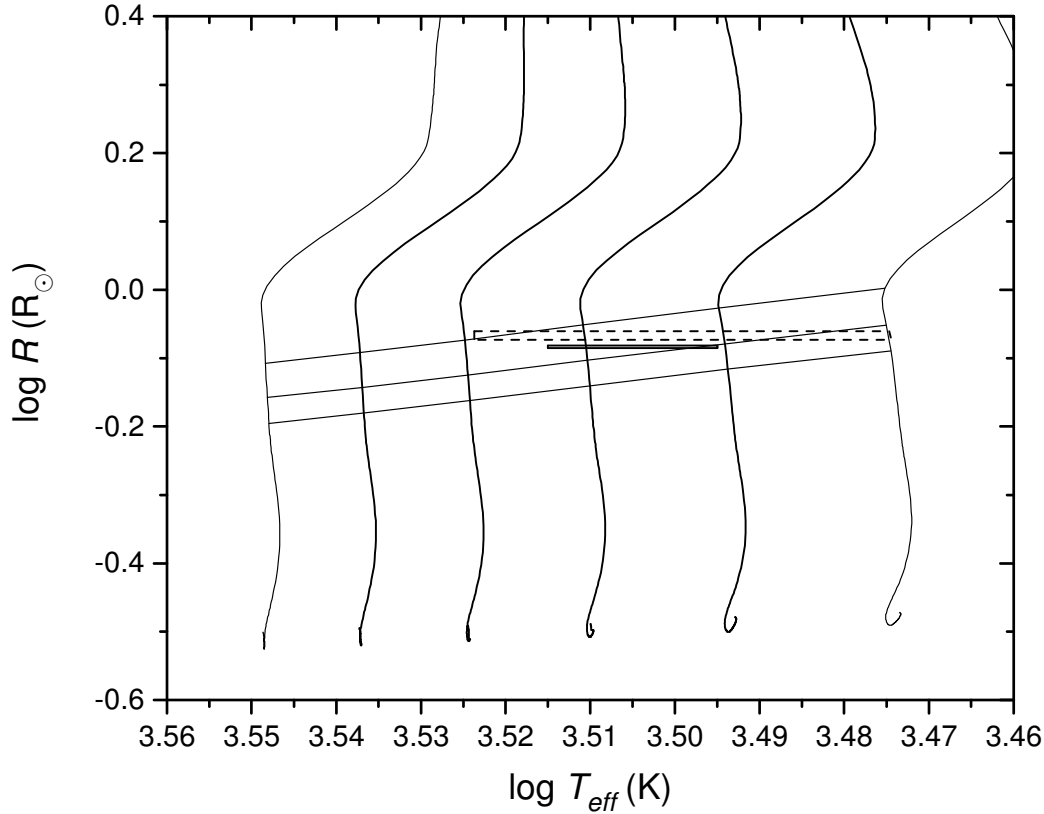

Figure 3. Analogous to Fig. 1, except for models from set C (see Section 4.1) for M = 0.32 $M_\odot$. Here, the spot coverage parameter ranges from 0 (leftmost track: "standard evolution") to 0.5 (rightmost track: heavily spotted) in increments of 0.1.

5.2 The CC system in USco

In Figure 4, we show pairs of evolutionary tracks in the $R - T_{eff}$ diagram which are obtained for models of mass 0.12 $M_\odot$ (red lines) and 0.11 $M_\odot$ (blue lines) when we use boundary conditions belonging to set A (see Section 4.1). The long rectangular boxes show the empirical locations of E2037 A (red) and E2037 B (blue). Also shown are isochrones. If we were to attempt to interpret the empirical values of the stellar radii in terms of non-magnetic models (as plotted on the leftmost pair of tracks), the empirical radii of the two components would require ages of 9.03 ± 0.54 Myr for E2037 A and 7.35 ± 0.53 Myr for E2037 B. According to this interpretation, the two components of the CC system cannot be coeval: they differ in age by about $2\sigma$. An additional difficulty with the nonmagnetic model interpretation is that the models would predict effective temperatures for both components that are ~300 K hotter than the empirical values.

However, when we consider magnetic models, the ones which fit the empirical values best (by passing closest to the centers of the rectangular boxes) are found to have ages of 12.8 ± 1.4 and 12.6 ± 1.6 Myr for components A and B, respectively. These ages are consistent with coevality for the components. As regards the best fit models, the magnetic inhibition parameters are found to have the values $\delta$ = 0.027 ± 0.006 and 0.038 ± 0.006 in components A and B respectively. Combining these values of the best-fitting $\delta$ with the surface gas pressure in the CC models (according to eq. 1.3), the corresponding surface vertical field strengths are found to be 340 – 430 G in component A and 420 – 500 G in component B. Why does the field turn out to be stronger in component B? The reason is that the oversizing of component B is definitely more extreme than in component A. To see this, note that if standard evolution of coeval components were to be applicable, component B should be smaller than component A. Specifically, at a particular age (e.g. 8 Myr), the red isochrone in Fig. 4 intersects the leftmost red track at a radius of log $R$ = -0.333, while the blue isochrone *at the same age* intersects the leftmost blue track at log $R$ = -0.365. Thus, standard evolution predicts that component B should have a radius which is 7-8% *smaller* than the radius of component A. (Similar quantitative differences are also found if we examine where the isochrones for 11 and 14 Myr intersect the leftmost tracks.) However, in fact, the empirical radii (David et al. 2016) show that component B is *not* smaller than component A. Instead, the opposite is the case: component B is found to have a radius which is 7-8% *larger* than component A. Thus, relative to the "standard" evolutionary tracks, the empirical oversizing of component B is more extreme than the oversizing of component A. This has the effect that, when we attempt to interpret the empirical data in terms of a magnetic model, a stronger magnetic field is required to fit the empirical radius in component B.

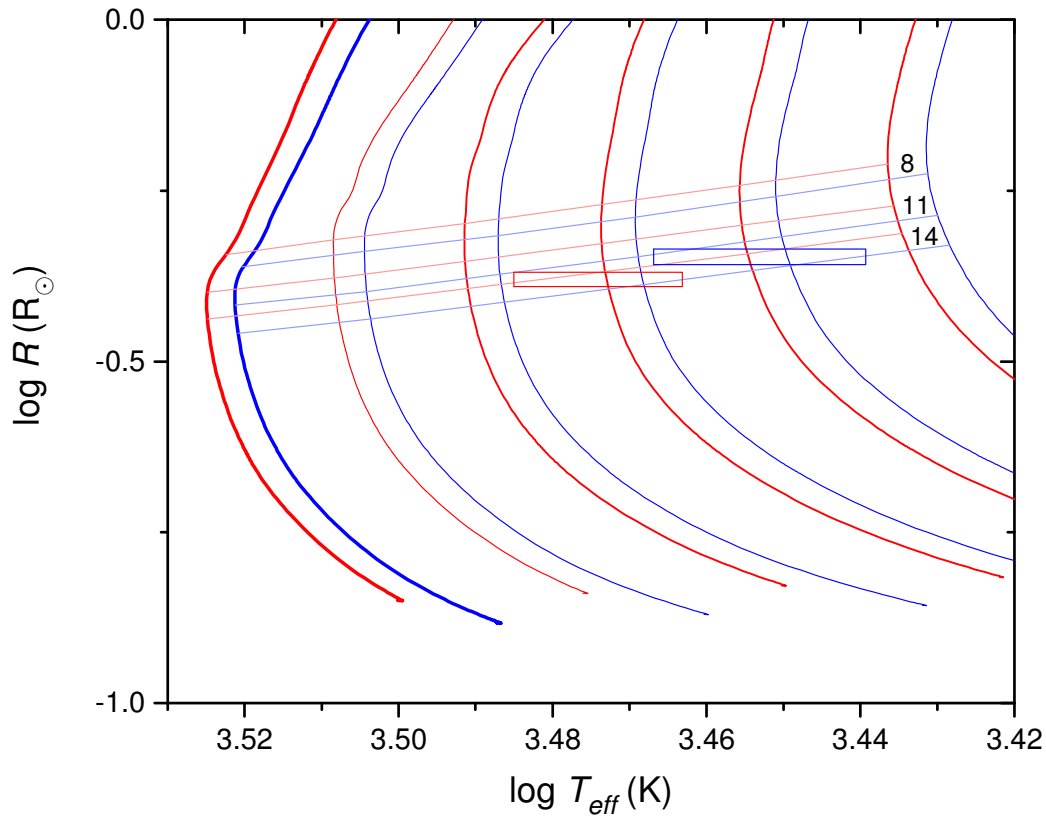

Figure 4. Comparison of the locations of the CC stars in USco (E2037A and B) in the $R - T_{eff}$ diagram with models from set A (see Section 4.1) for $M$ = 0.12 $M_\odot$ (red) and 0.11 $M_\odot$ (blue). The magnetic inhibition parameter ranges from 0 (leftmost tracks: nonmagnetic "standard evolution") to 0.05 (rightmost tracks) in increments of 0.01. The rectangles show the positions of component A (red) and B (blue) using the results of David et al. (2016). Also shown are isochrones for both masses at ages 8, 11 and 14 Myr.

Figure 5 is analogous to Figure 4 except the models are from set B. Because of the smaller convective efficiency in set B models (resulting from a shorter mixing length $\alpha=1.0$, rather than $\alpha=1.7$ in set A), the radii in set B models, even in the non-magnetic limit, are already systematically larger than those in set A models. The difference in radii can be quantified by comparing the location of the intersection of (say) the 8 Myr isochrones with the left-most red track: in Fig. 4, this occurs at log $R$ = -0.333, while in Fig. 5, it occurs at log $R$ = -0.316. That is, the nonmagnetic model from set B has a radius which exceeds the equivalent set A model by 4%. As a result, smaller fields suffice to give the oversizing that is needed to fit the empirical parameters. From the magnetic models, we find ages of 13.2 ± 1.5 and 13.0 ± 1.6 Myr for E2037 A and E2037 B, respectively. Once again, we note the advantage of using magnetic models: the components are found to be coeval. The values found for the magnetic inhibition parameter are $\delta$ = 0.020 ± 0.006 and 0.032 ± 0.007. The corresponding surface vertical field strengths are 280 – 390 G and 380 – 470 G.

Figure 6 is analogous to Figure 4, except the models are from set C. The spot models give ages of 13.0 ± 1.5 and 13.1 ± 1.7 Myr for E2037 A and E2037 B, respectively. The values found for the spot fraction parameter are $f_s$ = 0.22 ± 0.07 and 0.36 ± 0.07. As was noted above for the RZ system, there is no obvious way to predict the surface magnetic field strength which corresponds to a given fractional areal spot coverage.

Also as regards the CC system in USco, we draw attention to a similar quantitative conclusion that can be drawn from Figs 4-6 as was drawn for the RZ system shown in Figs. 1-3. Whether we use boundary conditions from Sets A, B, or C, the ages which emerge from our best-fit magnetic models of the CC stars in USco5 are found to lie in the range 12.6 ± 1.6 to 13.2 ± 1.5 Myr. These results are consistent with the ages obtained for the more massive stars in USco: 11 ± 3 Myr (as described in Section 1.2 above).

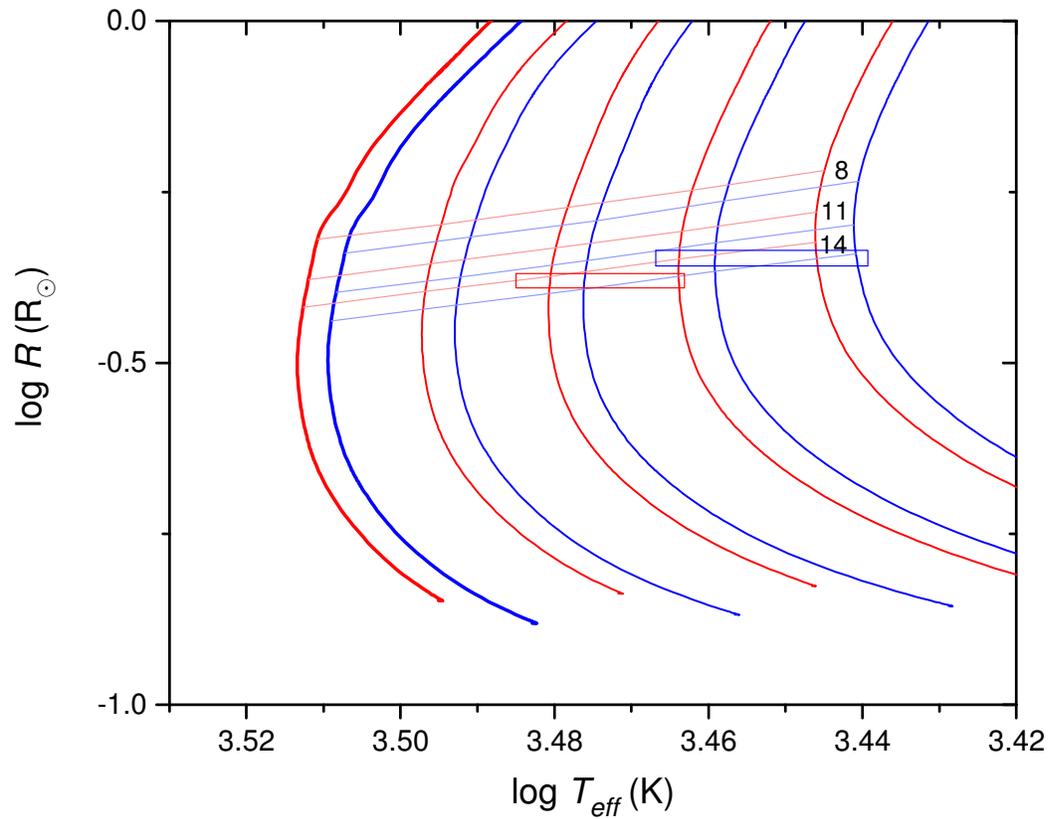

Figure 5. Analogous to Figure 4 except the models plotted here are from set B (see Section 4.1). The magnetic inhibition parameter ranges from 0 (leftmost tracks: nonmagnetic "standard evolution") to 0.04 (rightmost tracks) in increments of 0.01.

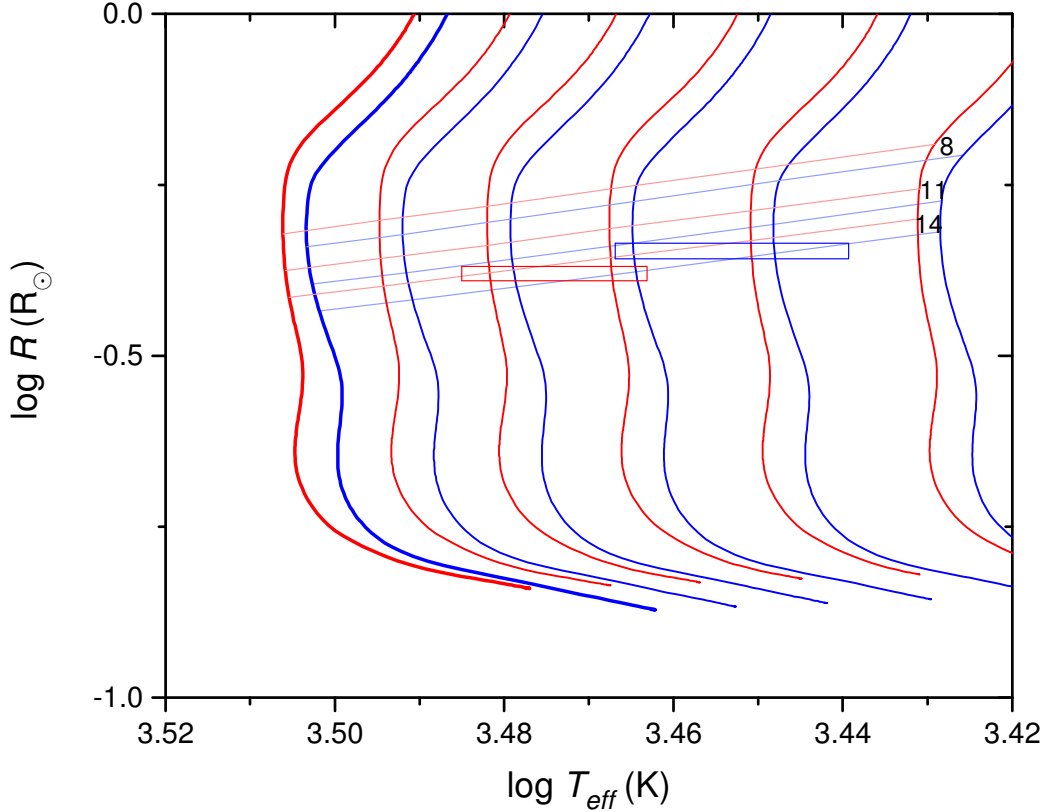

Figure 6. Analogous to Figure 4, except that the models plotted here are from set C (see Section 4.1). The spot coverage parameter ranges from 0 (leftmost track: unspotted "standard evolution") to 0.5 (rightmost track) in increments of 0.1.

5.3. The magnetic field ceiling and its role

Our choice of magnetic field profile has the vertical component of the field, $B_v$, increasing inwards proportional to $\sqrt{P_{gas}}$ until we arrive at a radial location where $B_v$ reaches a specified value: we refer to this specified value as the ceiling field, $B_{ceil}$. The mass coordinate $m_c$ of the position where $B_v$ first reaches the value $B_{ceil}$ (as we move inward from the surface of the star) will be illustrated for a variety of $B_{ceil}$ values in Fig. 8 below. Interior to $m_c$, our models assume that the magnetic field retains the same strength, $B_{ceil}$, all the way to the center of the star, thereby causing the value of $\delta$ to decrease to much smaller values as we approach the center of the star. For the models described so far in this paper, we have specified $B_{ceil}$ = 1 MG.

The question is: is there any way to check whether or not $B_{ceil}$ = 1 MG is a reasonable choice for the maximum field strength inside one of our stars? And as a corollary question, if it could be demonstrated that our choice of $B_{ceil}$ is implausible, how much would our conclusions change if we were to choose different values for $B_{ceil}$ in our USco stars?

At first sight, it might appear that there is little support for a value of $B_{ceil}$ as large as 1 MG in a star of mass 0.3 $M_\odot$ such as we consider here. For example, Yadav et al. (2015) have modelled the field generation by a distributed dynamo in a slowly rotating fully convective star, and have found that a small seed field grows exponentially until it reaches a state of rough equipartition between magnetic and kinetic energy densities. They find that the peak magnetic field strength is ~13 kG and occurs in the interior at radial location ~0.3 $R_*$ for a model of 0.3 $M_\odot$. In support of this conclusion of Yadav et al., we have found that, in our model of mass 0.32 $M_\odot$ on the early main sequence, the maximum field determined from equipartition with the energy density of convective motions would indeed be of order 5 – 10 kG. If these "convective equipartition" field strengths were to apply to our USco stars, then our choice of 1 MG for $B_{ceil}$ would be too large by a factor of 100.

We suggest that fields of order 10 kG (which are readily obtained in completely convective stars: see Yadav et al. 2015) may be regarded as setting a *lower limit* on the value of $B_{ceil}$ in the low-mass stars we are considering in USco. However, because of rotational effects, the possibility of generating stronger fields is worth considering. We estimate that, in our models of USco5, the maximum field strength could be as large as ~1 MG provided the rotation periods are of order 2 d.

To the extent that these suggestions are valid, we now turn to quantifying how the results obtained so far in this paper would be altered if we allow $B_{ceil}$ to take on different values. In an earlier paper (MacDonald & Mullan 2014), we have already discussed, in a context of older stars, how the value of $B_{ceil}$ affects the derived values of the surface field strength, $B_{surf}$. In our 2014 paper, we modeled the components of the double-line eclipsing binaries, YY Gem, CM Dra and CU Cnc. Based on independent age constraints, the stars in the first two binary systems are old enough to be in the main sequence phase of evolution. [The evolutionary state of the stars in the CU Cnc system is not so clear. Ribas (2003) suggested from it space velocity that, like YY Gem, CU Cnc might be a member of the Castor moving group, which indicates that the stars are close to the zero-age main sequence. However, CU Cnc has been determined to have heavy element abundance [Fe/H] ~ 0.3 (Gaidos et al. 2014; Terrien et al. 2015) significantly higher than that of Castor, [Fe/H] = 0.1 (Gebran et al. 2016) and association of these stars seems unlikely.]

Our 2014 modelling showed that, for main sequence (or near main sequence stars), the derived $B_{surf}$ values are remarkably insensitive to the choice of $B_{ceil}$ value. Specifically, in that earlier work, we found that $B_{surf}$ differed by no more than a factor of 2 even when we allowed $B_{ceil}$ to take on values which ranged over 4 orders of magnitude (from $10^4$ to $10^8$ G). If we were to approximate this behavior by a power law, $B_{surf} \sim B_{ceil}^a$, the exponent $a$ would be negative, with an absolute value no larger than 0.08. Now, in the case of USco stars, we can report that we find similar insensitivity to the choice of $B_{ceil}$. This can be seen from figure 7 which shows how $B_{surf}$ depends on $B_{ceil}$ for 0.32 $M_\odot$ set A models which match the mean of the David et al. (2016) measurements of the radii of the components of USco5 at age 11 Myr, and also for 0.11 $M_\odot$ set A models which match the David et al. (2016) measurement of the radius of E2037 B at age 11 Myr. The results in Fig. 7 indicate that if we reduce $B_{ceil}$ by a factor of 100 from 1 MG to 10 kG, but retain the constraint that the models must in all cases replicate the empirical radius of the star which is being modelled, we find that the values of $B_{surf}$ in our USco stars increase by a factor of only 1.35-1.4, indicating a power law with exponent $a$ = - 0.07. The small absolute value of $a$ is striking: the surface field strengths in our magneto-convective models are only very weakly affected by changes in our choice of $B_{ceil}$.

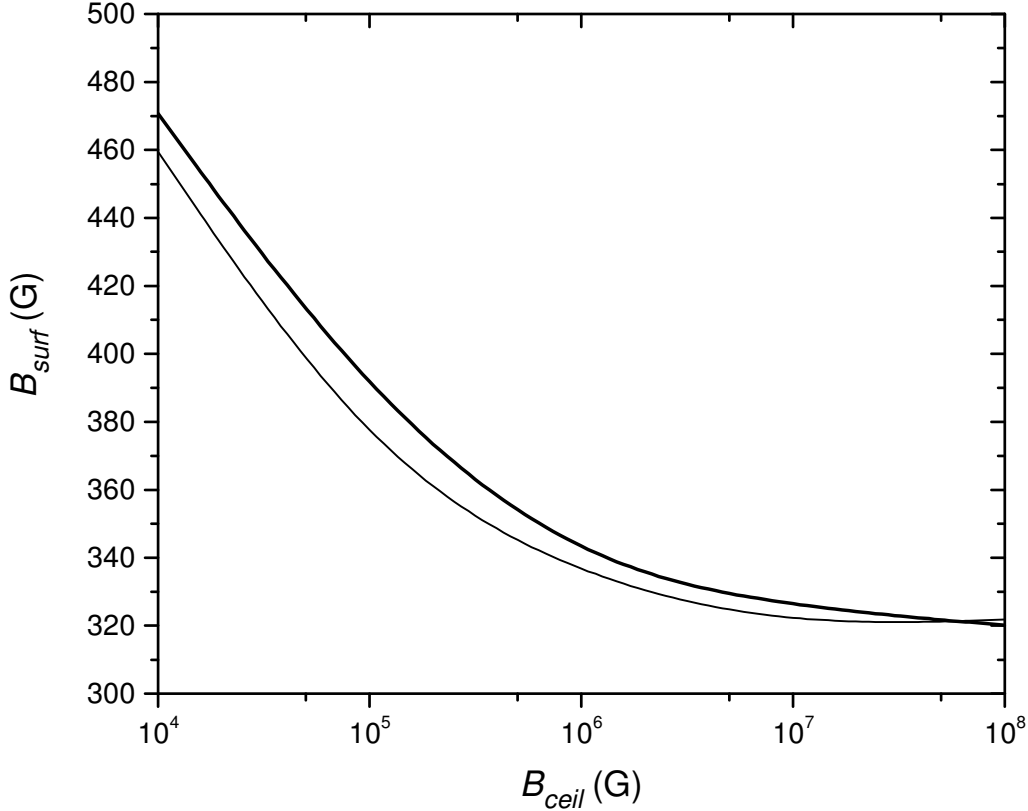

Figure 7. Dependence of predicted surface field strength on adopted value of the magnetic field ceiling for set A models of USco5 (thick line) and E2037 B (thin line).

At fixed age and stellar radius, we also find that the effective temperature is insensitive to the choice of $B_{ceil}$. For the USco5 models, the difference in $T_{eff}$ for $B_{ceil}$ values of 10 kG and 1 MG is only 40 K. This difference in $T_{eff}$ is small compared to the observational uncertainty in $T_{eff}$. Therefore, we can report that changing the $B_{ceil}$ value by factors of 100 does not significantly affect our main conclusion that inclusion of magnetic effects restores coevality for the stars in USco.

To reveal the regions of the star where the magnetic effects are largest in our magneto-convective models, we plot in figure 8 the relative difference in density between our non-magnetic model and 4 magnetic models of USco5 which were computed using different values of the ceiling field strength. We stress that all of the models shown in Fig. 8 have identical radii, namely, the empirical radius reported by David et al (2016): in order to achieve this in the presence of differing ceiling fields, the value of $\delta$ differs also from one model to the next. In Figure 8, the stellar surface is on the left. The short lines indicate the depth where the magnetic field strength first reaches its ceiling value as we move inward from the surface. We see that the greatest relative density change occurs deeper than where the ceiling is first reached. We also see that the structural effects of the field are not localized but occur over most of the star. For example, for the $B_{ceil}$ = 10 kG case, the greatest relative difference in density, $\Delta \log \rho = -0.35$, occurs at mass depth $2 \times 10^{-6}$ $M_\odot$, but even at the stellar center, $\Delta \log \rho = -0.29$. As expected, the depth at which the greatest relative density difference occurs does increase with the value of $B_{ceil}$.

We stress here that the main effect of the field in these models is to suppress convective energy transport. Even if the field is imposed only in the outer parts of the star in a region containing a relatively small amount of mass, e.g. ~ $10^{-7}$ M$_\odot$ in the $B_{ceil}$ = 10 kG case, suppression of convective energy transport would lead to a damming up of heat in the whole of the interior with the consequential result of causing the star to expand globally.

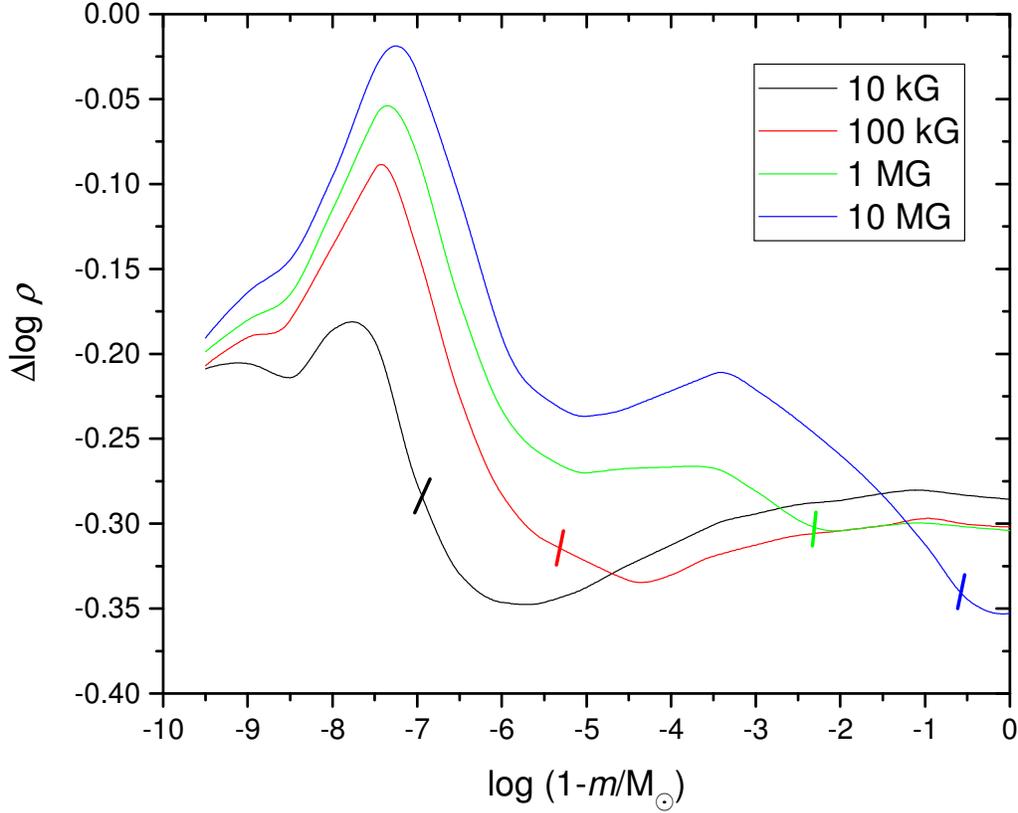

Figure 8. Relative difference in density between a magnetic model and the non-magnetic model plotted against mass depth in units of stellar mass. The key refers to the magnetic field ceiling value, $B_{ceil}$. The short lines show the depth at which the field first reaches $B_{ceil}$.

6. DISCUSSION

6.1. Best-fit estimates of "age(s)" of USco stars: coevality restored

According to the 3 distinct sets (Sets A, B, C) of magnetic/spotted evolutionary tracks which we have computed, we find that the best-fitting models (according to the most recent empirical data: David et al. 2016) to the components of the RZ system in USco have ages as follows: 9.9 ± 2.0 Myr and 10.6 ± 2.0 Myr (Set A), 10.2 ± 2.0 Myr and 10.8 ± 2.0 Myr (Set B), and 9.6 ± 2.1 Myr and 10.4 ± 2.2 Myr (Set C). Note that the age determinations are insensitive to the mixing length ratio, because of the degeneracy in changing mixing length ratio and changing the magnetic inhibition parameter or spot fraction (see MacDonald & Mullan 2014). Our models suggest that the best-fit age for the RZ stars in USco is in the range 9.6-10.8 Myr, with error bars of about 2 Myr.

We also find that the best-fitting models for the components of the CC system in USco have ages as follows: 12.8 ± 1.4 and 12.6 ± 1.6 Myr (Set A), 13.2 ± 1.5 and 13.0 ± 1.6 Myr (Set B), and 13.0 ± 1.5 and 13.1 ± 1.7 Myr (Set C). Our models suggest that the best-fit age for the CC stars is in the range 12.6 - 13.2 Myr, with error bars of about 1.7 Myr.

If we assume *a priori* that the 4 low-mass stars we have studied in USco are co-eval and provide independent USco age estimates, then the most likely age is 11.8 Myr (set A), 12.1 Myr (set B) or 12.0 Myr (set C). The standard deviation is 0.9 Myr, independent of which model set is used. These age estimates from our magnetic models of 4 low-mass stars in USco overlap very well with the age determination of 11 ± 3 Myr reported for more massive stars in USco (with spectral types B, A, G, as well as Antares) by Pecaut et al. (2012). As a result, the "problem" of non-coevality in USco is resolved by our magnetic models.

6.2. Predictions of surface magnetic field strengths

Our models indicate that the strengths of the vertical component of the magnetic field at the stellar surface in the case of our best fit models (assuming $B_{ceil}$ = 1MG) for USco range from 230 to 470 G for the two RZ stars, and from 280 to 500 G for the two CC stars. If instead of a ceiling field of 1 MG, the maximum field strength in our models is assumed to be only 10 kG, then the above surface field strengths should be increased by a factor of 1.35 – 1.4 (see Figure 7): in such cases, the surface field strengths could be as large as 600 – 700 G. Surface fields of several hundred Gauss have previously been estimated when we applied our magneto-convective model to three low-mass solar neighborhood binaries YY Gem, CM Dra and CU Cnc (MacDonald & Mullan 2014). Although the stars in the latter 3 systems are certainly not as young as the stars in USco, nevertheless the results are pertinent to the present discussion because the masses of the various components once again span the range of masses from RZ to CC. Thus, at 0.21 and 0.23 $M_\odot$, the components of CM Dra are definitely in the CC class, whereas at 0.60 $M_\odot$, each component in YY Gem is definitely in the RZ class. The components of CU Cnc, at 0.40 and 0.43 $M_\odot$, are also in the RZ class.

In contrast to the approach we adopt here to model magneto-convection in stars, Feiden & Chaboyer (2012, 2013) have used a different approach to modelling the effects of magnetic fields on stellar structure and evolution. Feiden & Chaboyer (2013) report that a total surface magnetic field strength of ~4 kG is required to replicate the empirical radii of the components of YY Gem. When MacDonald & Mullan (2014) applied their model to YY Gem, they found that the vertical component of the surface magnetic field is required to be in the range 320 – 350 G.

Recently, Feiden (2016) has applied the Feiden-Chaboyer model in order to address the age discrepancy in USco between high-mass and low-mass stars. Feiden's derived a consistent age of 10 Myr for high- and low-mass stars in USco. In order to obtain this age for the low-mass stars, Feiden did not allow the surface field strength in a model of a low mass star in USco to be a floating parameter to be determined by fitting magnetic evolutionary tracks to the empirical data. Instead, in what he describes as a "critical" choice, Feiden prescribed the surface field strength to be equal *a priori* to the equipartition field strength $B_{eq}$ in the gas at optical depth $\tau = 1$. For stars with a range of masses, this leads to surface fields of 0.76 kG in a 1.7 $M_\odot$ star, and 2.64 kG in a 0.1 $M_\odot$ star at an age of 10 Myr. Thus, for the low-mass stars we have discussed here, Feiden *assumes* a surface field of 2.44 – 2.64 kG in order to calculate his magnetic models of low-mass stars in USco. These strengths are larger by a factor of up to 10 compared to our own estimates of 230 – 700 G on the same stars.

At first sight, our results for surface field strengths in USco stars do not appear to be consistent with those of Feiden. But this is not necessarily true, because of a significant difference between the model approaches: our model predicts the *vertical* component of the field, whereas Feiden's model deals with the *total* magnetic field strength. The fact that our field estimates are *smaller* than Feiden's indicates that *both* results could be true. (If our field estimates had turned out to be larger than Feiden's, then one (or both) of the results would have to be considered unacceptable.) Let us now consider if observational tests might help resolve the inconsistency.

6.3. Observational tests?

Observations of field properties in low mass stars have progressed greatly in recent years. But different observations give different information about the fields.

Polarimetric observations of the Stokes V parameter in the context of Zeeman-Doppler imaging (ZDI) allow observers to reconstruct in principle the three components of the magnetic field vector on a stellar surface in spherical coordinates, i.e. radial, azimuthal, and meridional (Donati et al. 2003). However, because of cancellation of field components when one integrates across the stellar disk, Stokes V measurements cannot effectively detect strong fields in small active regions. On the other hand, non-polarimetric observations of the Stokes parameter I allow estimates of the total field strength (e.g. Reiners & Basri 2009). As a result, the field $B_V$ which is reported by a Stokes V measurement is always weaker than the field $B_I$ reported from a Stokes I measurement. For example, Morin et al. (2010) report that in a sample of late M dwarfs, $B_V/B_I$ is found to have a value which never exceeds 0.2, and is in some cases less than 0.05.

In order to compare the results of our modeling with the observations, it is important to note the following. The field for which our magneto-convective models yield information (once the magnetic inhibition parameter $\delta$ has been determined) is the vertical component $B_v$ (see eq. 1.3). It can be argued (Mullan, MacDonald & Townsend 2007) that, among the components of the stellar magnetic field which can be derived from ZDI (toroidal, meridional, poloidal), the one which is most closely related to $B_v$ is the poloidal field at higher latitudes. Therefore, we are especially interested in observational results which provide information about the vector components, rather than the total strength of the field. In view of this, the difference between Feiden's approach and ours may be considered as follows: our models yield fields which should be compared with $B_V$ data, whereas Feiden's models yield fields which should be compared with $B_I$ data. The fact that our values for surface field strengths in USco stars are smaller than Feiden's values by a factor which may be as large as 10 is consistent with the range of $B_V/B_I = 0.05 - 0.2$ reported by Morin et al. (2010).

For a more specific observational comparison between our results and polarimetric data on young stars, we refer to results by Folsom et al. (2016): the youngest stars in their sample were estimated to be 20 Myr old, i.e. within a factor of 2 of our USco stars. Using ZDI data, Folsom et al. found that the peak field strength $B_{peak}$ on the surfaces of 15 young solar-type stars was in the range 40 – 800 G. For each star, ZDI data also allowed Folsom et al. to extract the fraction $\phi$ of field energy which is in the poloidal component $B_p$. This allows us to calculate the field component which is of most interest to us: $B_p = B_{peak}\sqrt{\phi}$. Inserting data for the 10 stars in Folsom et al's sample with the strongest fields, we find $B_p = 160 - 620$ G. To the extent that our estimates of $B_v$ on the stellar surfaces of USco stars (230-700 G) can be equated to the value of $B_p$ (as suggested by Mullan et al. 2007), we see that the poloidal field strengths in our 4 USco stars are encouragingly consistent with the results which have been obtained observationally by Folsom et al. (2016) for stars with ages that are almost as young as those of USco.

7. CONCLUSION

The USco association has, in the past, been notable for claims that the ages of the high-mass stars are apparently much older than those of the low-mass stars. We have obtained magnetic models for 4 low-mass components of two binary systems which belong to the young USco association. One of the binary systems contains stars with masses in the range 0.3 – 0.35 $M_\odot$: such stars may contain a radiative zone during certain evolutionary stages. The second binary system contains stars which are in the range 0.10 – 0.12 $M_\odot$: such stars are expected to be completely convective in USco. The magnetic models that we have obtained are successful in fitting the empirical data on $R$ and $T_{eff}$ at ages which (within the error bars) are identical for all 4 low-mass stars. Significantly, the ages which we have found to provide the best fit to the low-mass stars are also consistent with previous (independent) estimates of the ages of the more massive stars in the association. As a result, all of the stars in the USco association can now be regarded as coeval provided that magnetic fields are present in (at least) the low-mass stars. (Fields in higher-mass stars are not expected to lead to significant alterations to standard evolution.)

Although we would like to compare our theoretical results for the vertical component of the fields on the surfaces of USco stars with empirical information, no data are yet available which would allow us to make a direct comparison. In order to make a meaningful comparison, ZDI observations of the low-mass USco stars would be very welcome.

ACKNOWLEDGEMENTS

We thank the anonymous referee for constructive criticism. This work is supported in part by NASA/Delaware Space Grant.